# Driving Condition-Aware Multi-Agent Integrated Power and Thermal Management for Hybrid Electric Vehicles

Hanghang Cui [1], Arash Khalatbarisoltani [1], Jie Han [2], Wenxue Liu [1], Muhammad Saeed [3], and Xiaosong Hu [1]

[1] Department of Mechanical and Vehicle Engineering, Chongqing University, Chongqing, China,
[2] Department of Aeronautical and Automotive Engineering, Loughborough University, Loughborough, England
[3] Department of Chemical Engineering, Shanghai Jiao Tong University, Shanghai, China

**Abstract:** Effective co-optimization of energy management strategy (EMS) and thermal management (TM) is crucial for optimizing fuel efficiency in hybrid electric vehicles (HEVs). Driving conditions significantly influence the performance of both EMS and TM in HEVs. This study presents a novel driving condition-aware integrated thermal and energy management (ITEM) framework. In this context, after analyzing and segmenting driving data into micro-trips, two primary features (average speed and maximum acceleration) are measured. Using the K-means approach, the micro-trips are clustered into three main groups. Finally, a deep neural network is employed to develop a real-time driving recognition model. An ITEM is then developed based on multi-agent deep reinforcement learning (DRL), leveraging the proposed real-time driving recognition model. The primary objectives are to improve the fuel economy and reduce TM power consumption while maintaining a pleasant cabin temperature for passengers. Our simulation results illustrate the effectiveness of the suggested framework and the positive impact of recognizing driving conditions on ITEM, improving fuel economy by 16.14% and reducing TM power consumption by 8.22% compared to the benchmark strategy.

## 1-Introduction

Developing new efficient energy vehicles is an important initiative to strengthen the transportation sector to tackle increasing concerns about global warming and reduce emissions. There are multiple candidates like electric vehicles (EVs), hybrid electric vehicles (HEVs), and fuel cell vehicles (FCVs) [1]. However, HEVs are seen as one of the most suitable options for transitioning to fully green transportation alongside

EVs and FCVs. Energy management strategies (EMSs) have been developed to manage power allocation among the multiple power sources and can be categorized primarily into three groups: rule-based [2], optimization-based [3-7], and learning-based [8-10]. In addition to the EMS, thermal management (TM) of the engine, battery, and cabin has a critical impact that cannot be ignored. Although EMS and TM problems may seem like two independent challenges, in terms of fuel savings and components' lifetime, both EMS and TMS are, in fact, highly coupled. Temperature significantly affects the components and TM's power consumption, usually provided by battery, thereby affecting overall HEV power consumption, battery charge sustainability, and battery pack lifetime. Moreover, maintaining cabin, battery pack, and engine temperatures within their optimal ranges while minimizing fuel consumption and battery degradation is challenging. For instance, [11] presents a health-conscious predictive EMS and investigates the impact of battery electro-thermal-aging models on the EMS results. In another study, an EMS is suggested that considers the preheating of the battery packs in low-temperature driving scenarios [12]. To address the multi-physical and complex dynamics of the EMS and TM of the cabin, battery, and engine, integrated thermal and energy management (ITEM) is proposed in the literature as a valuable opportunity [13-15].

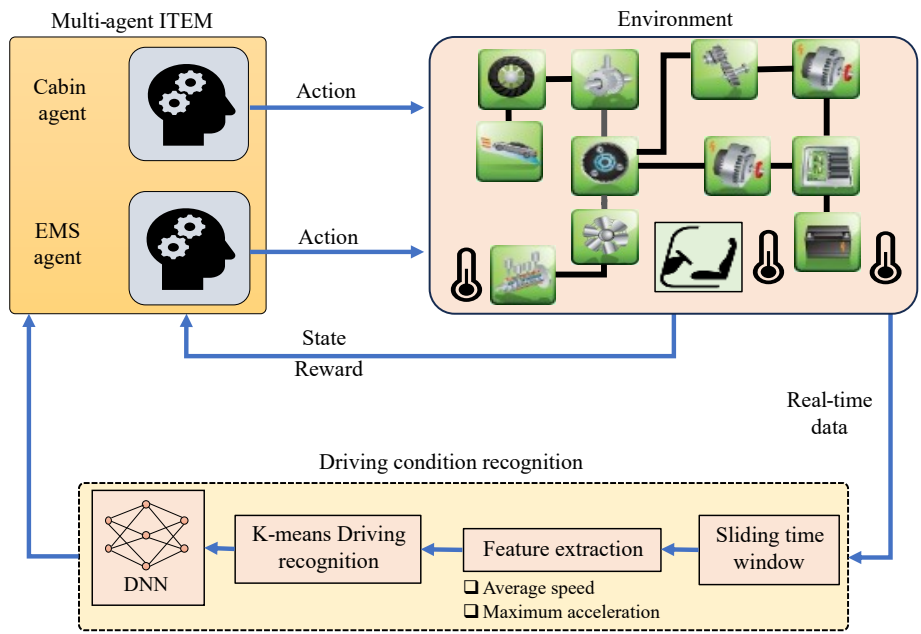

Fig.1. Driving-style-aware multi-agent ITEM

Driving conditions also play a crucial role in the performance of EMSs and TMs, and neglecting these conditions can lead to a decline in the effectiveness of EMS strategies in real-world driving scenarios. Driving cycles can be categorized based on various factors, including vehicle speed, acceleration, and jerk, using both unsupervised learning methods (such as K-means, fuzzy C-means, and kernel density estimation) and supervised learning methods (Bayesian estimation, K-nearest neighbors, and support vector machines). Several studies have integrated driving conditions with EMSs, including adaptive equivalent consumption minimization strategy, stochastic model predictive control, and deep reinforcement learning (DRL).

Even though there have been successful investigations, the majority of previous strategies have focused on driving-aware EMSs while mostly overlooking the impact of driving conditions on TM performance, TM power consumption, and cabin temperature. To this end, this study proposes a novel driving-style-aware ITEM strategy, with specific attention to exploring the impacts on fuel economy, TM power consumption, and cabin comfort. Taking into account the recognized driving conditions, as shown in Fig. 1, a multi-agent DRL-based ITEM is proposed to simultaneously find the optimal power allocation policy and learn the impact of driving conditions on ITEM to improve fuel economy and cabin comfort temperature. Extensive simulation results validate the proposed driving condition-aware multi-agent DRL approach, demonstrating significant improvements in fuel economy and TM power consumption while keeping cabin temperatures within a comfortable range, compared to the benchmark strategy.

The remainder of this study is structured as follows: The power-split HEV powertrain modelling is detailed in Section 2. The driving condition recognition algorithm is presented in Section 3. Section 4 describes the driving style-aware DRL ITEM. Section 5 discussed the simulation results, followed by the conclusion.

**2-HEV Powertrain Modelling**

The power-split HEV model, including electrical, mechanical, and thermal power loops as displayed in Fig. 2, provided by Autonomie, a simulation tool created by the Argonne National Laboratory. In the powertrain, the first motor $M_1$ is connected to the torque coupler and is responsible for propulsion purposes and regenerative braking. The second motor $M_2$ is coupled directly to the ICE via a planetary gearbox, functioning as a generator and assisting in starting the ICE. The HEV parameters used in this study are outlined in Table I. The driving force of the HEV $F_a$ can be calculated by

$$F_d = F_w + F_f + F_z + F_a, \tag{1}$$

$$F_w = \frac{1}{2}\rho C_{d A_f} v^2, \tag{2}$$

$$F_f = m\, gf cos\theta, \tag{3}$$

$$F_z = m\, g\, sin\, \theta, \tag{4}$$

$$F_a = m\, a \tag{5}$$

where $F_d$ represents the HEV driving force, $F_w$ and $F_f$ are the air and friction force resistances, respectively. $F_z$ and $F_a$ denote the slope resistance and the requested driving force for acceleration, respectively, $\rho$ represents the air density, $C_d$ is the air resistance coefficient, $A_f$ is the HEV front area, $m$ represents the vehicle's total mass, $g$ is the gravitational acceleration, $f$ is the frictional resistance coefficient, $\theta$ is the road slope, and $a$ is vehicle acceleration. The requested power $P_d$ can be formulated by

$$P_d = F_d\, v, \tag{6}$$

where v is the vehicle velocity. The demanded torque T_ps and power P_ps from the power-split system can be calculated by

$$T_{ps} = \frac{F_d\, r_w}{r_d(\eta_w \eta_d)^{sgn(P_d)}}, \tag{7}$$

$$P_{ps} = \frac{P_d}{(\eta_{w,i}\eta_d)^{sgn(P_d)}}, \tag{8}$$

where $r_w$ denotes the wheel radius, $r_d$ represents the total drivetrain gear ratio, $\eta_w$ is the wheel efficiency, and $\eta_d$ stands for drivetrain efficiency. The power-split device's quasi-static equations can be expressed by

$$\frac{r_g T_e}{1+r_g} = -T_{R_g} = -r_g T_{m_2}, \tag{9}$$

$$(1 + r_g)\omega_e = r_g \omega_{R_g} + \omega_{m_2}, \tag{10}$$

$$T_{R_c} = r_t T_{m_1},$$ (11)

$$\omega_{m_1} = r_t \omega_{R_c},$$ (12)

where $T_{R_g}$ and $\omega_{R_g}$ denote the torque and speed of the ring gears of the gearbox, while $T_{R_c}$ and $\omega_{R_c}$ represent the torque and speed for the ring gears of the torque coupler, respectively.

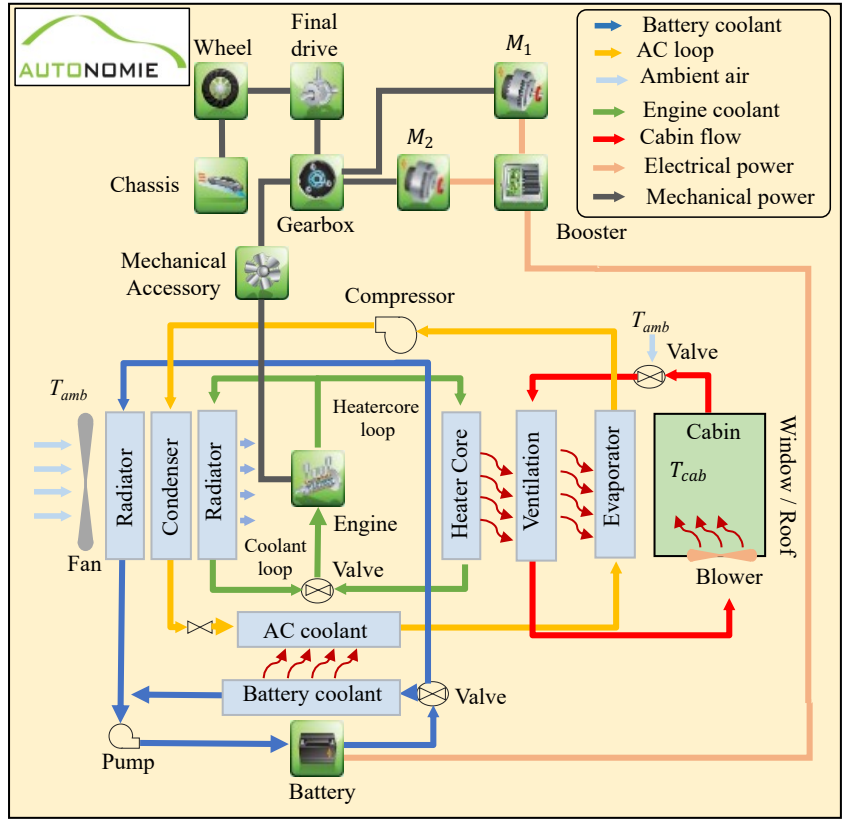

Fig. 2. Schematic of electric, mechanical, and thermal power flows of the power-split HEV

$T_e$ and $\omega_e$ are the torque and speed for the $ICE$, respectively, the torque and speed of $M_1$ and $M_2$ are presented by $T_{m_1}$ and $\omega_{m_1}$ as well as $T_{m_2}$ and $\omega_{m_2}$, respectively. The gear ratios for the planetary gearbox and the torque coupler are $r_g$ and $r_t$, respectively. Finally, the power-split device interaction with the wheels can be expressed by

$$\omega_{R_g} = \omega_{R_c} = \frac{r_d v}{r_w}.$$ (13)

The battery power $P_b$ is calculated by

$$P_{bat} = P_{m_{1,dc}} + P_{m_{2,dc}} + P_{bat_{aux}},$$ (14)

where $P_{m_{1,dc}}$ and $P_{m_{2,dc}}$ represent the power of $M_1$ and $M_2$, respectively, and $P_{bat_{aux}}$ is the power of the auxiliary system provided by the battery pack. The battery cell output power $P_{cell}$ is formulated by

$$P_{cell} = \frac{P_{bat}}{n_{bat}}$$ (15)

where $n_{bat}$ is the cell numbers of the battery pack. The battery is modeled as an first-order RC equivalent circuit as formulated by

$$\begin{cases} \dot{V}_1 = -\dfrac{1}{R_1 C_1} V_1 + \dfrac{1}{C_1} I_{cell} \\ V_t = V_{ocv} - V_1 - I_{cell} R_0 \end{cases}$$ (16)

where $V_{ocv}$ presents the open circuit voltage, $V_t$ denotes the terminal voltage, $V_1$ stands for the polarization voltage, $R_0$ is the ohmic resistance, $R_1$ is the polarization resistance, $C_1$ denotes the polarization capacitances, and $I_{cell}$ is the battery cell current. All of the battery parameters depend on temperature and state of charge (SOC). Interested readers can find details about the significance and impact of battery parameters in [16-18]. $I_c$ can be obtained from

$$I_c = \frac{(V_{oc} - V_1) - \sqrt{(V_{oc} - V_1)^2 - 4R_0 P_c}}{2R_0}$$ (17)

The battery cell SOC changes can be obtained by

$$\Delta SOC = \frac{-I_c}{3600 \, (Q_n - \Delta Q)},$$ (18)

where $I_c$ represents the battery cell current, $Q_n$ is the nominal cell capacity, and $\Delta Q$ represents the cycling capacity degradation, which can be calculated using the aging model. The battery cell temperature is formulated by [19, 20]

$$\dot{T}_{bat} = -\frac{h_c A_c}{m_{bat}\, c_p} T_{bat} + \frac{1}{m_{bat}\, c_p}\left(Q_g - Q_a - Q_c\right) + \frac{h_c A_c}{m_{bat}\, c_p} T_{amb}, \qquad (19)$$

where $T_b$ represents the cell bulk temperature, $h_c$ is for the convection coefficient, $A_c$ denotes the convection area of the battery cell, $T_{amb}$ is the ambient temperature, $m_{bat}$ is the cell mass, $c_p$ is the specific heat capacity of the cell, $Q_g$ represents the heat generation, $Q_a$ is the heat rejected by the air-cooling system through fan operation, and $Q_c$ is the heat rejected by coolant system. A semi-empirical aging formula is employed to model the capacity degradation of the battery cell. The battery thermal model parameters are $c_p = 1350$ J/(kg°C), $h_c = 5$ W/(m$^2$°C), $m_b = 40$ J/°C, and time constant is 5 s. The capacity loss $Q_l$ is quantified as a percentage reduction from the initial battery capacity at the start of its lifespan.

$$Q_l = M \exp\left(-\frac{E_a}{T_b\, R_g}\right) A_d{}^z, \qquad (20)$$

where $Q_l$ denotes the percentage of capacity loss of the battery, and $M$ is the pre-exponential factor, which is a function of the C-rate. $T_{bat}$ and $R_g$ represent the lumped battery temperature and the ideal gas constant, respectively. $E_a$ denotes the cell activation energy, $A_d$ represents the discharged ampere-hour throughput, and $z$ is the calibration constant. The total discharged throughput can be calculated using

$$A_{tol} = [20/M \exp\left(-\frac{E_a}{R_g T_{bat}}\right)]^{\frac{1}{z}}. \qquad (21)$$

The number of cycles until the battery end-of-life (EOL), denoted as $N$, is achieved by

$$N = \frac{3600 A_{tol}}{Q_n}. \qquad (22)$$

The state-of-health (SOH) change is calculated by

$$SOH = 1 - \frac{|I_c|}{2N Q_n}. \qquad (23)$$

The engine's coolant temperature $T_{cl}$ is formulated by

$$\dot{T}_{cl} = \frac{1}{C_s M_e} (\dot{Q}_f - \dot{Q}_w - \dot{Q}_{ex} - \dot{Q}_{con} - \dot{Q}_{rad} - \dot{Q}_{cab} + \dot{Q}_{hs}), \quad (24)$$

where $C_s$ and $M_e$ represent the equivalent specific heat capacity and the mass of the $ICE$, respectively. $\dot{Q}_f$ is the heat generated by fuel (gasoline), $\dot{Q}_w$ represents the heat converted to work, $\dot{Q}_{ex}$ and $\dot{Q}_{con}$ symbolize the heat passed through with exhaust gas and the heat rejected by air convection, respectively. $\dot{Q}_{rad}$ is the heat rejected by coolant, $\dot{Q}_{cab}$ represents the heat transferred to the cabin, and $\dot{Q}_{hs}$ represents the heat added by heat storage. The cabin temperature $T_{cab}$ is modeled by [21]

$$\dot{T}_{cab} = \frac{1}{M_{cab} C_{cab}} (\dot{Q}_{hvac} + \dot{Q}_{sun} + \dot{Q}_{roof} + \dot{Q}_{win} + \dot{Q}_t), \quad (25)$$

where $M_{cab}$ and $C_{cab}$ denote the equivalent air mass and heat capacity of the cabin, respectively. $\dot{Q}_{hvac}$ is the heat flow from the climate control system, $\dot{Q}_{sun}$ represents the radiation heat from the sun and $\dot{Q}_{roof}$ and $\dot{Q}_{win}$ are the heat flows by convection from the roof and window, respectively. $\dot{Q}_t$ is the heat load by the heat transmission.

### 3-Real-Time K-Means Driving-Recognition

This study proposes a real-time driving recognition model based on the K-means algorithm and deep neural network (DNN). Ten standard driving profiles are designated; among them, nine driving cycles are used for initial clustering, and one is considered the unseen profile. A fixed timestep is selected to divide the driving dataset, which includes the chosen nine standard driving cycles, into 505 micro driving cycles, each lasting 20 seconds. After the driving division step, two driving characteristic parameters (average speed and maximum acceleration) are selected to characterize each micro trip. The main reason for choosing only a limited number of features is that the direct clustering of the resulting characteristic parameter matrix does not lead to high data dimensions or a heavy computational burden. To determine the labels for each driving cycle, a K-means cluster analysis approach is utilized to cluster the micro trips into three categories.

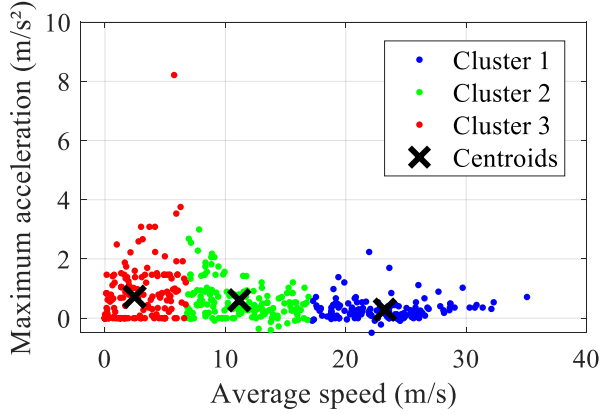

Fig. 3. K-means clustering algorithm categorized the driving data into three primary clusters

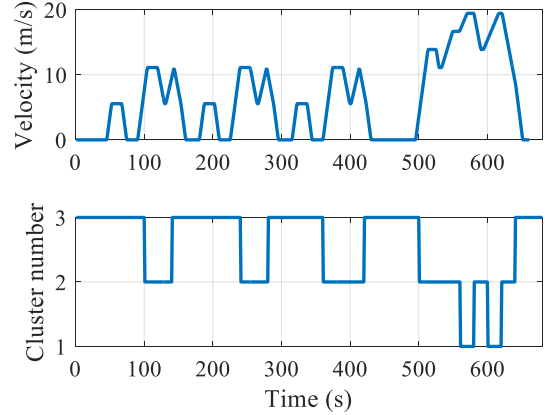

Fig. 4. Clustering results of the JN1015 standard driving cycle

The model first calculates the distance between the current input sample and each cluster center. Then, it classifies them according to the principle of minimum distance to obtain the cluster label, which indicates the recognized type of driving cycle or style. Fig. 3 presents the K-means clustering results. For better illustration, the cluster number for the JN1015 standard driving cycle is shown in Fig. 4.

After obtaining the labeled micro trips using the K-means algorithm, a DNN was employed to build an instantaneous driving recognition model. The two features derived from the original driving data and the K-means clustering output are used as inputs to the DNN. The trained real-time DNN-based driving recognition model achieves high accuracy, around 90%, compared to the micro trips that are clustered solely using K-means.

**4-Driving-Aware Energy and Thermal Management Formulation**

The DDPG algorithm is selected to efficiently learn the optimal ITEM actions. The proposed control strategy is composed of two DRL agents, cabin TM and EMS, which aim collaboratively to reduce fuel and TM units' power consumption while keeping cabin temperature as close as possible to a comfortable range. The state space of the cabin agent is proposed as follows:

$$s_{Cab} = \{ T_{Cab}, e_{Cab}, V_s, V_a, DC \} \tag{26}$$

where $T_{\text{Cab}}$ represents the cabin temperature, $e_{\text{Cab}}$ denotes the deviation of the cabin temperature $T_{\text{Cab}}$ from the cabin target temperature $T_{\text{Cab,t}} = 22\ °C$, $V_s$ and $V_a$ represent the vehicle's longitudinal speed and acceleration, respectively, and $DS$ stands for the recognized driving condition, which is obtained from the previously discussed model. In the developed driving-aware ITEM, accurate SOC estimation is crucial, as it significantly impacts both fuel consumption and the battery lifetime of HEVs. An effective SOC estimation approach with low computational time is particularly valuable, as it enables real-time provision of necessary information for optimal energy management [22, 23]. Several low-computational and highly accurate approaches have been investigated in the literature to address this [24-26]. The action space of the cabin agent, denoted as $A_{Cab}$, is presented as follows

$$A_{Cab} = \{\, a_h, a_{AC}\} \tag{27}$$

where $a_h = a_{ac} = \{0, 1\}$ denotes the cabin heater and AC system operation modes, respectively. The cabin DRL agent supervises two proportional-integral-derivative controllers that manage the cabin AC system and heater system. The reward function for the cabin TM agent, denoted as $r_{Cab}$, is designed as

$$r_{Cab} = -\alpha_1 e_{Cab}{}^2 - \alpha_2 P_{Cab}, \tag{28}$$

where $e_{\text{Cab}}$ represents the deviation in the cabin temperature $T_{\text{Cab}}$ from the target temperature $T_{\text{Cab,t}} = 22\ °C$, $P_{Cab}$ denotes the cabin climate control system power consumption, and $\alpha_{1,2}$ are the cabin agent's reward function coefficients. The state spaces of the EMS agent, denoted as $s_{EMS}$, is designed as follows

$$s_{EMS} = \{V_s, V_a, \Delta SoC, a_{last}, DC\} \tag{29}$$

where $\Delta SoC$ is the SoC deviation from the initial level and $a_{last}$ is the last action taken by the EMS agent. The action space of the EMS agent, denoted as $A_{EMS}$, selects the increase or decrease value of the ICE output power, as described by

$$A_{EMS} = \begin{cases} a_1 = 10kW, a_2 = 5kW, a_3 = 1kW, \\ a_4 = 500\ W, a_5 = 0, a_6 = -1kW, \\ a_7 = -5kW, a_8 = \text{Engine off} \end{cases}. \tag{30}$$

The ICE output power is considered to operate only along the optimal working line. The reward function for the EMS agent, denoted as $r_{EMS}$, consists of the instantaneous fuel consumption of the ICE and the penalty term for sustaining battery charge, formulated by

$$r_{EMS} = -\beta_1 \dot{m}_f - \beta_2 (SoC - 0.7)^2 \tag{31}$$

where $\dot{m}_f$ represents the fuel consumption rate and the last term is included to sustain the $SoC$ level, and $\beta_{1,2}$ are coefficients.

## 5-Results and Discussion

This section aims to assess the performance of the driving-aware ITEM. Ten standard driving profiles are considered in this investigation; among them, nine driving cycles are for the initial clustering and training of the DRL agent, and one is considered an unseen profile. Figure 5 displays the ITEM framework with and without the driving recognition information comparison results under the NYCC driving cycle and shows that driving-aware ITEM effectively keeps the final SOC around 0.65, which is very close to the initial value. By leveraging the real-time DNN driving recognition model, the suggested framework reduces fuel consumption by 16.14% compared to the version without driving recognition. Moreover, the cabin temperature is more stable than in the DRL without driving input, which infers that the passenger feels a more comfortable temperature based on the proposed driving conditions-aware ITEM strategy. In addition, the power consumption of the TM shows an 8.22% reduction compared with the strategy without driving recognition inputs. These results highlight that integrating the driving recognition model with ITEM successfully enhances the performance of both EMS and TM in HEVs.

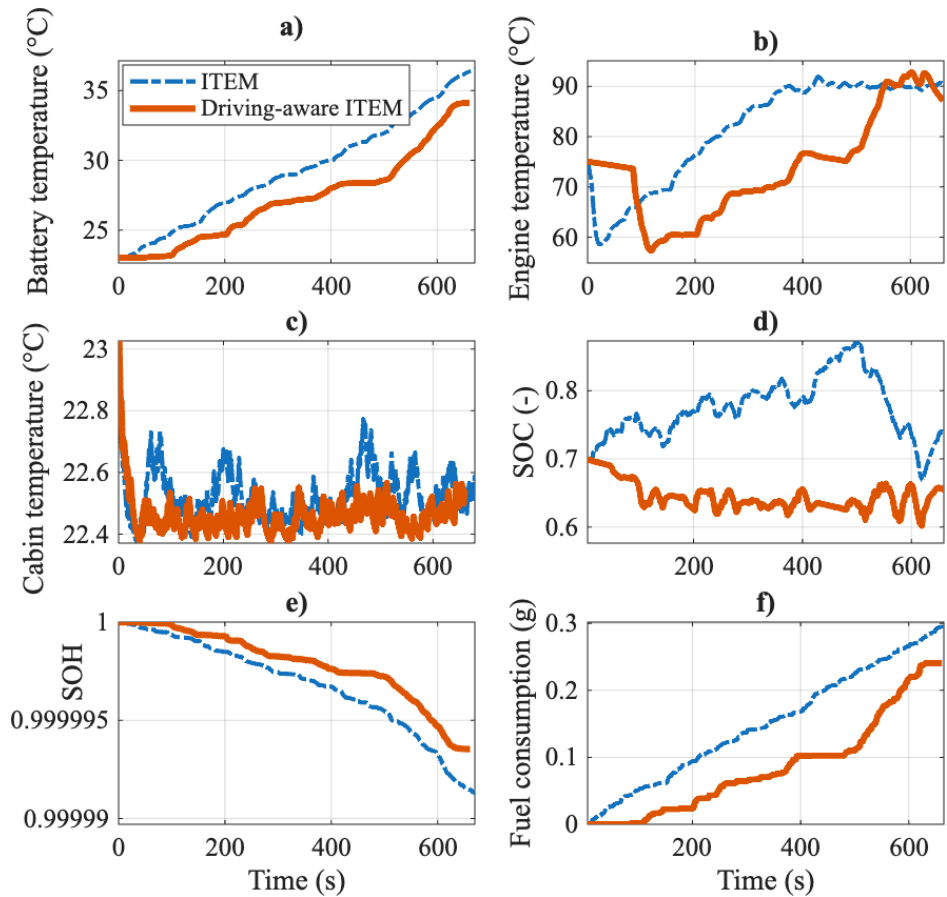

**Fig. 5.** Comparison of the results under NYCC: a) battery temperature, b) engine temperature, c) cabin temperature, d) SOC, e) SOH, f) fuel consumption

**Conclusion**

This study proposes a real-time driving-aware ITEM for HEVs. The driving recognition model is developed based on the K-means algorithm and the training of a DNN model. The co-optimization of the EMS and TM is formulated within the multi-agent DRL framework, with the main objectives of improving fuel economy and reducing TM's power consumption while ensuring that cabin temperature remains within a comfortable range. In order to integrate the driving recognition model, the results are added as an extra input to the ITEM framework. The results highlight that considering the real-time driving recognition model for ITEM can decrease fuel consumption and TM power consumption while keeping the cabin in a comfortable range compared with the standard ITEM without driving information. The driving-aware ITEM strategy delivers notable improvements: fuel consumption is reduced by 16.14%, and TM power consumption by 8.22%, compared with ITEM methods without driving recognition information while keeping the cabin in a comfortable range.